\documentclass[letter]{aa}
\usepackage{graphicx,txfonts,natbib}

\begin{document}
\title{Folding ion rays in comet C/2004 Q2 (Machholz)\\and the connection with the solar wind\thanks{based on observations collected with the Flemish 1.2m Mercator telescope at the Roque de los Muchachos observatory, La Palma, Spain, and with the SOHO and ACE satellites}}
   \subtitle{}
   \author{P. Degroote\inst{1}
          \and
           D. Bodewits\inst{2,3}
          \and
           M. Reyniers\inst{1}\fnmsep\thanks{Postdoctoral fellow of the Fund for Scientific Research, Flanders}
           }
   \offprints{P. Degroote}
   \institute{Instituut voor Sterrenkunde, Departement Natuurkunde en Sterrenkunde, K.U.Leuven, Celestijnenlaan 200D, 3001 Leuven, Belgium\\
             \email{pieter.degroote@ster.kuleuven.be\ }\email{maarten.reyniers@ster.kuleuven.be}
         \and
              KVI atomic physics, University of Groningen, Zernikelaan 25, 9747AA Groningen, The Netherlands\\
              \email{bodewits@kvi.nl}
         \and
              NASA Postdoctoral Fellow, Goddard Space Flight Center, Solar System Exploration Division, Mailstop 693, Greenbelt, MD 20771, USA
             \email{dennis.bodewits@ssedmail.gsfc.nasa.gov}}
   \date{Received October 23, 2007; accepted November 26, 2007}
   \authorrunning{P. Degroote et al.}
   \titlerunning{Folding ion rays in C/2004 Q2 (Machholz)}

  \abstract
  {}
  {The appearance of folding ion rays in cometary comae is still not very well understood, so our aim is to gain more insight into the role of the local solar wind in the formation of these structures.}
  {Comet C/2004 Q2 (Machholz) was intensively monitored during its closest approach to Earth (January 2005) with the CCD camera Merope mounted on the Flemish 1.2m Mercator telescope, in three different bands (Geneva U and B and Cousins I). Spectacular ion rays, thin ionic structures rapidly folding tailward, were recorded in the U band during one night, January 12th.}
  {Data from the SOHO satellite that was extrapolated corotationally to the position of the comet showed that the ion rays were formed during a sudden change in the in-situ solar wind state. We were able to succesfully correlate a high-speed solar wind stream with the appearance of folding ion rays.}
  {To our knowledge, this is the first clear observational evidence that folding ion rays in cometary comae are produced by a sudden change in the local solar wind state.}
  \keywords{Comets: individual: C/2004 Q2 (Machholz), Comets: general, Sun: solar wind}

  \maketitle
%

\section{Introduction}

\begin{figure*}
\includegraphics[width=\textwidth]{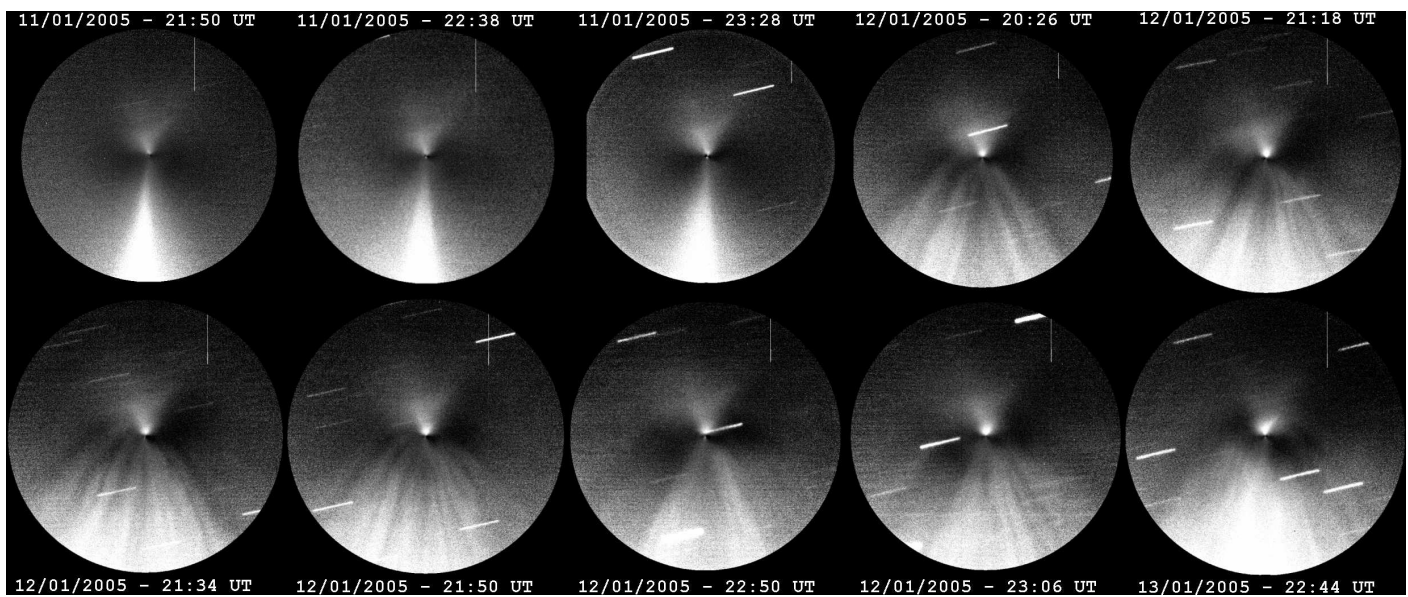}
\caption{Azimuthal renormalised frames of 6 of 8 observations in the U band of comet C/2004 Q2 (Machholz) on 2005 Jan. 12, together with three observations of the night before (upper left), and one on the night after (lower right). The trails on the images are due to background stars and indicate the comet's motion. All images taken on the 12th show clear ion rays, while on the 11th, only the typical sharp ion tail is visible; no ion rays appear. In the right lowerhand corner, it is clear that the ion rays gradually disappear. All images have the same scale and color coding, the direction of the Sun is upwards, and the field of view radius is approximately 4.3\,$\times$10$^4$\,km.}\label{fig:pdegroote:foldings_comp}
\end{figure*}

A comet can be considered to consist of a small nucleus ($\sim$10\,km) surrounded by a dusty gas cloud, dubbed the coma ($\sim$10$^5$\,km). On even larger scales, the coma is embedded in the halo, extending up to $\sim$10$^7$\,km. Two other typical components of comets in the vicinity of the Sun are the dust tail, consisting of relatively large neutral grains, and the ion tail, made of ionized gas. Basically, the two tails are seperated because the outward solar radiation pressure is higher than the inward gravitational pull. \citet{bierman51} showed that radiation pressure alone could not be responsible for the outward force on the ions. In this way, the morphology of comets was one of the first proofs for the existence of the solar wind.

Thus, variations in the solar wind result in a variable ion tail. Conversely, through certain properties of the cometary plasma structures, it may be possible to deduce valuable information on the solar wind. For example, a recent survey of Chandra comet observations has demonstrated that X-rays provide a quantitative diagnostics of the interacting solar wind \citep{bodewits07}. Other properties that have been exploited are tail disconnection events (DEs); see for example \citet{nieder78}, \citet{snow04}, and the recent tail rip-off from comet 2/P Encke \citep{vourlidas07} and statistical studies of the aberration angle of the main tail axis \citep{brandt73}. This sharply contrasts with the study of cometary folding rays, as they appear closer to the nucleus and thus require a higher spatial resolution. Current theories suggest that folding rays are tailward extensions of ion pile-up regions (receding envelopes) on the sunward side of the nucleus, but it is still not well known what exactly causes this phenomenon. Possible explanations include external origins: some studies invoke the crossing of the comet with IMF (interplanetary magnetic field) sector boundaries, others denote the IMF as just a source mechanism and put more emphasis on the cometary ionosphere. Still other studies claim they have a completely internal origin and are dependent on the nucleus' ion production rate \citep[see][for a review]{ip04}. This letter supports the idea that the solar wind is largely responsible for the appearance for cometary folding rays, as we found a clear indication that a fast cold wind (originating from coronal holes) triggers tail rays.

We observed comet C/2004 Q2 (Machholz) while it encountered a change in the solar wind regime. Such observations are rare, as there are a lot of variables that have to be set: firstly, we must have accurate solar wind data and detailed images of the comet with an adequate spatial and temporal resolution. Secondly, the observed comet has to be relatively close to the Earth, preferably in or near the ecliptic, to be able to extrapolate solar wind information to the position of the comet (as the solar wind is highly variable in both time and place). Furthermore, a solar event has to occur during the observing run and the effects have to be detected. It is quite a challenge to prepare for such an observation run, as many of these variables are not predictable on a long term and thus constant monitoring of solar wind data is necessary. In this light, the passage of comet C/2004 Q2 (Machholz) in January 2005 was an exceptional event since all of these conditions were set. 

\section{Observations, reduction, and image enhancement}
We obtained about 181 CCD-observations of comet C/2004 Q2 (Machholz), taken in three different broadband filters, namely Geneva U (61 observations of 600\,sec integration), Geneva B (61 observations of 60\,sec), and Cousins I (59 observations of 60\,sec). The images were taken with the Merope camera mounted on the 1.2m Mercator telescope on La Palma, Spain, between the 7th and 15th of January 2005. The camera consists of a 2158\,$\times$\,2044 pixel CCD with a field of view of 6.\arcmin5\,$\times$\,6.\arcmin5 and a resolution of 0.\arcsec19\,pix$^{-1}$. This translates to $\sim$50\,km\,pix$^{-1}$ at the position of the comet. The reduction of the CCD images is described in detail in \citet{reyniers08}.

The overwhelming brightness of the coma makes it difficult to study relatively faint structures such as ion rays. For a spherically symmetric, optically thin coma with a stationary outflow of dust grains, the projected brightness profile in the plane of the sky equals a $1/r$ distribution, with $r$ the projected radius in the sky \citep{wallace58}. The inner coma ($r$\,$\lesssim$\,10$^4$\,km) is thus very bright in comparison with other features. The situation changes beyond the outer parts of the coma, where the overall density of the gas and dust becomes too small to reflect much light, and the accumulation of ions in certain directions now becomes relatively high. Even in the inner coma, these features are present in the signal, but the signal has to be enhanced to make them visible to the naked eye. Many techniques have been developed \citep[see e.g.][for an overview]{schleicher04}, but each technique suppresses certain features while bringing out others, so they have been carefully examined and tested. In general, these techniques include dividing and subtracting the frames by empirical or theoretical (sunward) surface brightness profiles.
The downside here is that we lose information on azimuthal structures like shells \citep{schulz91}.

Other techniques include subtracting/dividing with rotated/translated copies, azimuthal renormalisation around the nucleus, and filtering with different kernel sizes, such as the Larson-Slaughter filter \citep{larson92} and Larson-Sekanina filter \citep{sekanina84}. We chose to subtract an azimuthally renormalised profile via
\[I(r,\theta)=I_0(r,\theta)-\sum\frac{I(r)}{n(r)}\]
where $n(r)$ is the number of pixels on a distance $r$ from the nucleus. To avoid artifacts due to the presence of field stars, it is usually better to replace the average by a median. In our case, the technique of azimuthal renormalisation has been shown to be both the fastest and qualitatively best procedure to enhance highly radial features such as ion rays, because it produces virtually no artifacts.


\section{Folding ion rays}

If we take a look at our dataset, we see that the observations in the U-filter display a very variable behaviour in the tailward direction, while the images in the I-band do not seem to change at all. This is a clear indication that the I-filter focuses on the cometary dust, while the U-filter shows ionic structures. The U-band is dominated by fluorescent emission from molecules like $\mbox{CO}^+$ (3787 \AA), $\mbox{N}_2^+$ (3914 \AA), and $\mbox{OH}^+$ (3600 \AA) \citep{lutz93}. For example, on 11 January, the typical sharp ion tail of the comet is clearly visible. The most remarkable event is, however, the appearance of thin ion rays on 12 January. On that date, the observations differ significantly compared to the images taken on previous and consecutive days; before they are not visible, and afterwards, they seem to disappear gradually (Fig.\,\ref{fig:pdegroote:foldings_comp}). Further investigations of the frames reveal a very symmetric pattern for about all observations on 12 January. 
The ion rays are symmetric around the tail axis, although possibly small deviations cannot be excluded.

\begin{figure}
\includegraphics[width=\columnwidth]{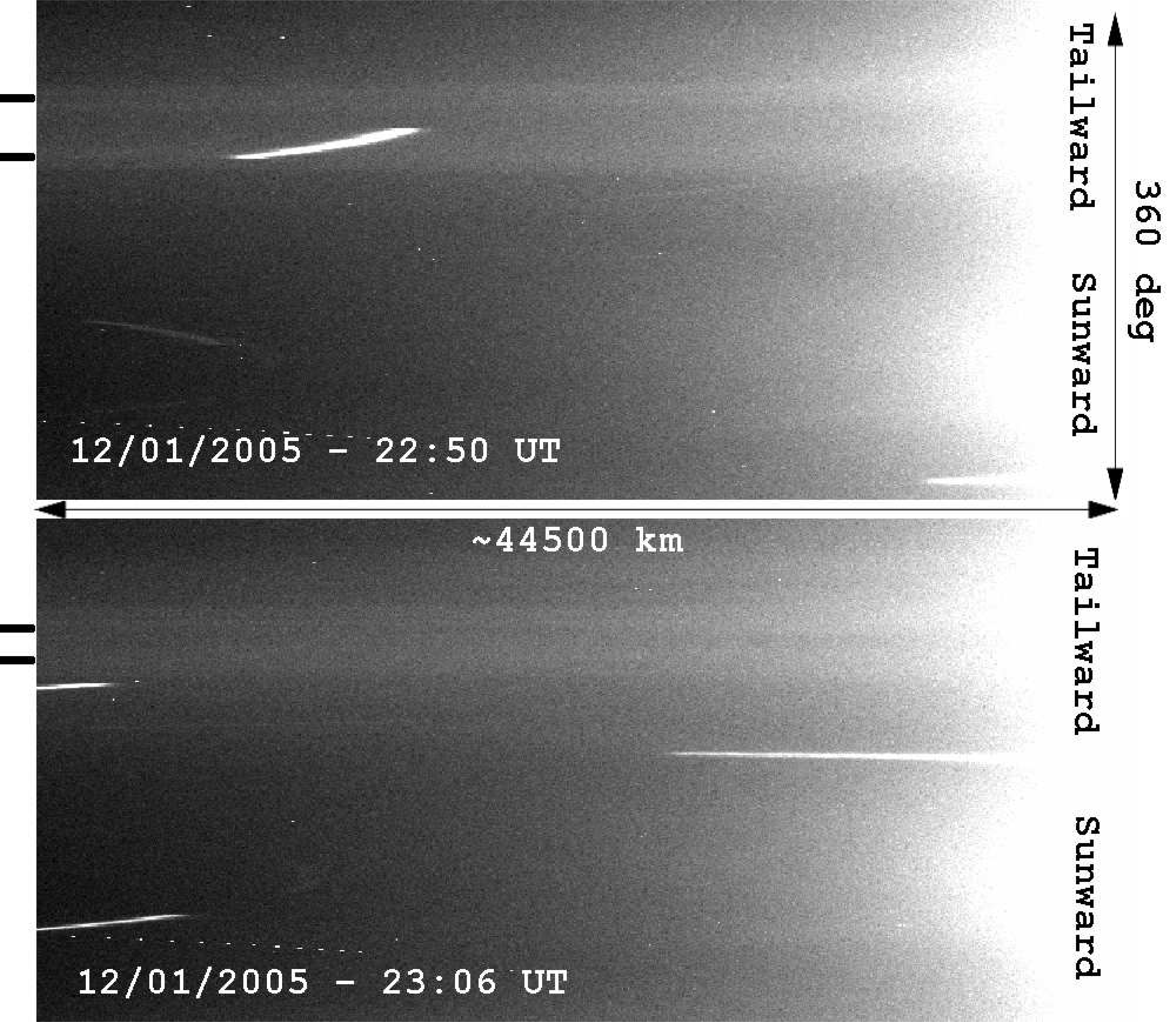}
\caption{Polar representation of images taken 2005 Jan. 12 at 22:50 and 23:06\,UT.  The comet's nucleus is to the right. In the upper section of both images, two ion rays are visible (indicated by tickmarks). These rays are folding along the tail's axis with an angular velocity of $\sim$\,0.9 degrees per minute.}\label{fig:pdegroote:foldings}
\end{figure}

On images taken at 22:50 UT and 23:06 UT (16 minute interval), the evolution of two ion rays can be traced. It is visually clear from their polar representation (Fig. \ref{fig:pdegroote:foldings}) the rays are folding along the tail axis of the comet. To derive an estimate of the angular velocity with which the rays converge, we integrated the flux radially over an interval of 70 pixels, starting at a distance of 530 pixels from the optocentre. This interval was chosen because of the high signal and the absence of stars. This way the signal was significantly enhanced, which made it possible to represent the position of the ion rays by the local maxima of a polynomial fit through the integrated flux as a function of the azimuthal coordinate. We concluded that these maxima folded around the tail axis with an angular velocity of $\sim$\,0.9 degrees per minute. To estimate the order of magnitude of the physical velocity associated to this value, we calculate that, at a distance where 1 pixel spans 1 angular degree (which is at about 3000 km from the nucleus), the velocity of the structure is $\sim$235\,km\,s$^{-1}$. This is too high to be of cometary origin, whereas it has the same magnitude as the solar wind speed. This brings us to a closer examination of the solar wind conditions at the position of the comet.

\section{Connection with solar wind through SOHO and ACE data}
SOHO and ACE are located at the first Lagrangian point L1 between the Sun and the Earth (ca. 0.01 AU from Earth). One of the many purposes of the SOHO satellite is to measure the solar wind velocity, which is done by the Proton Monitor on the CELIAS instrument \citep{hovestadt95}. Together with the measurements of the solar wind proton density taken by the same instrument, this gives us the solar wind flux as well.

\begin{figure}[htb]
\includegraphics[width=0.98\columnwidth]{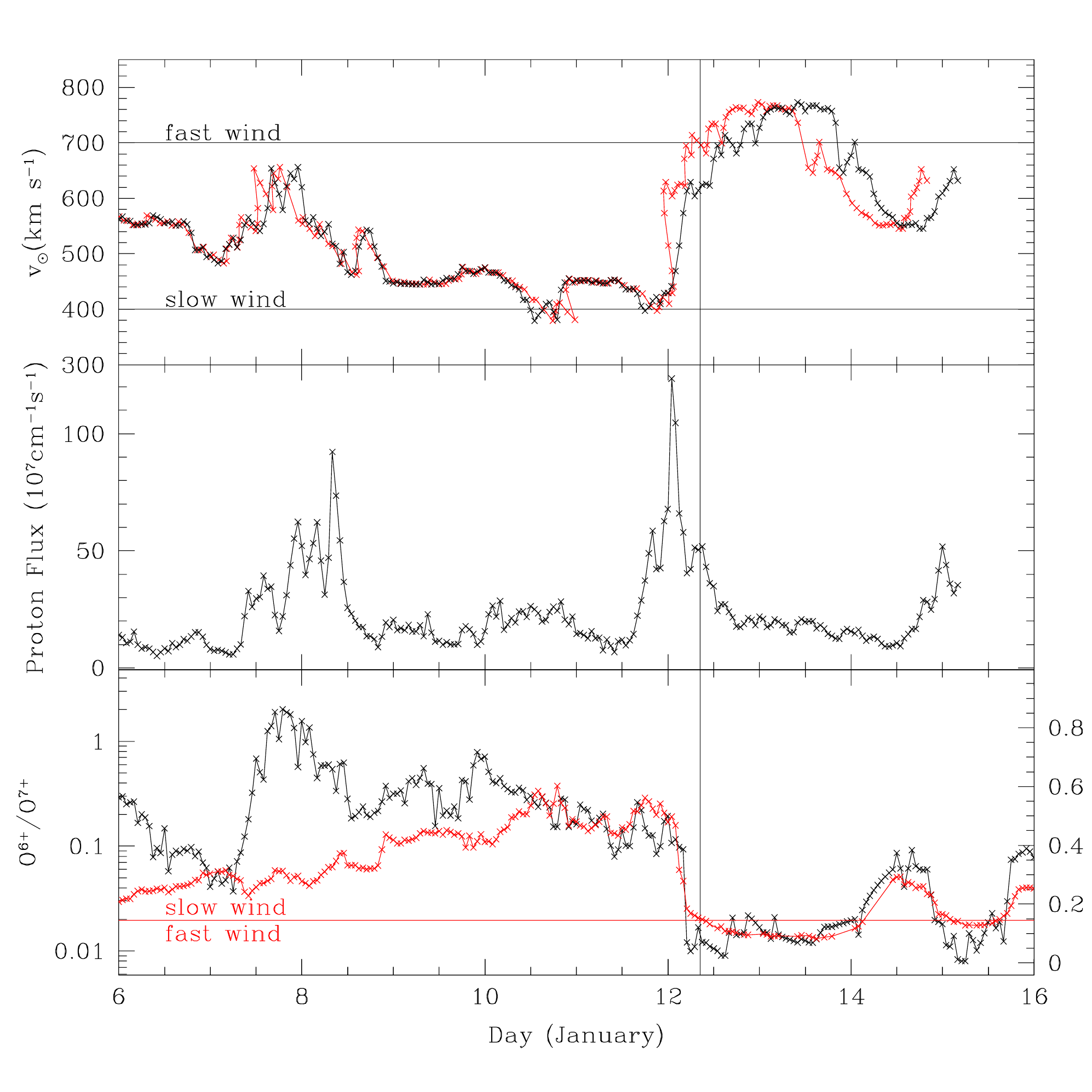}
\caption{{\em Upper panel:} Measurements of the solar wind velocity from SOHO/CELIAS (black line) and a corotational mapping (red line) at the location of C/2004 Q2 (Machholz). {\em Middle panel}: The proton flux from SOHO/CELIAS Proton Monitor at L1. The solar event on 12 January was due to a CIR, which is typically preceded by a sudden peak in the proton flux. {\em Lower panel}: Ratio of $\mbox{O}^{7+}/\mbox{O}^{6+}$ from ACE/SWICS. The red line represents the results from an empirically deduced formula \citep{zurbuchen06}, for which a low value (horizontal red line) indicates a fast, cold solar wind. The vertical black line denotes the first observation on 12 January.}\label{fig:pdegroote:soho_extrapol}
\end{figure}

From a first look at the SOHO data (Fig. \ref{fig:pdegroote:soho_extrapol}, {\em upper panel}), a clear discontinuity is seen in the solar wind velocity on 12 January at L1. We cannot immediately presume that this event also happened at the comet C/2004 Q2 (Machholz) position, because the solar wind has a finite velocity that is variable in both space and time as well. First of all there are two regimes of solar wind, namely the ``slow'' and ``fast'' regimes, with velocities around $400\mbox{ km s}^{-1}$ and $700\mbox{ km s}^{-1}$, respectively. Also, it is known that the solar wind can be very different at different latitudes. We can safely assume that the sudden rise in the wind velocity is due to a corotating interaction region (CIR) originating from a coronal hole, because there was a peak in the proton flux just before the wind velocity rose and stayed up for a day or two (Fig. \ref{fig:pdegroote:soho_extrapol}, {\em middle panel}), which is typical of a CIR \citep{nieder78}.

We used the measurements of the $\mbox{O}^{7+}/\mbox{O}^{6+}$-ratio from ACE/SWICS \citep{gloeckler98} to determine the nature of the solar wind streams (i.e. originating from coronal holes, coronal mass ejections (CME's), or simply slow solar wind). Oxygen was chosen because it is the most abundant heavy ion. The ratio of heavy highly-ionized ions $\mbox{O}^{7+}/\mbox{O}^{6+}$ gives us an idea of the temperature of the wind, therefore also of its origin. The low observed $\mbox{O}^{7+}/\mbox{O}^{6+}$ ratio implies a low-freeze in temperature, suggesting a cool origin on the sun. A ratio around $1\%$ is indeed in good agreement with values typical of a coronal driven wind \citep{zurbuchen06} (Fig. \ref{fig:pdegroote:soho_extrapol}).  Moreover, the SOHO/LASCO instrument detected no CME's heading for Earth at that time ({\ttfamily\small http://lasco-www.nrl.navy.mil/halocme.html}).

\begin{figure}[htb]
\centering\includegraphics[width=0.53\columnwidth]{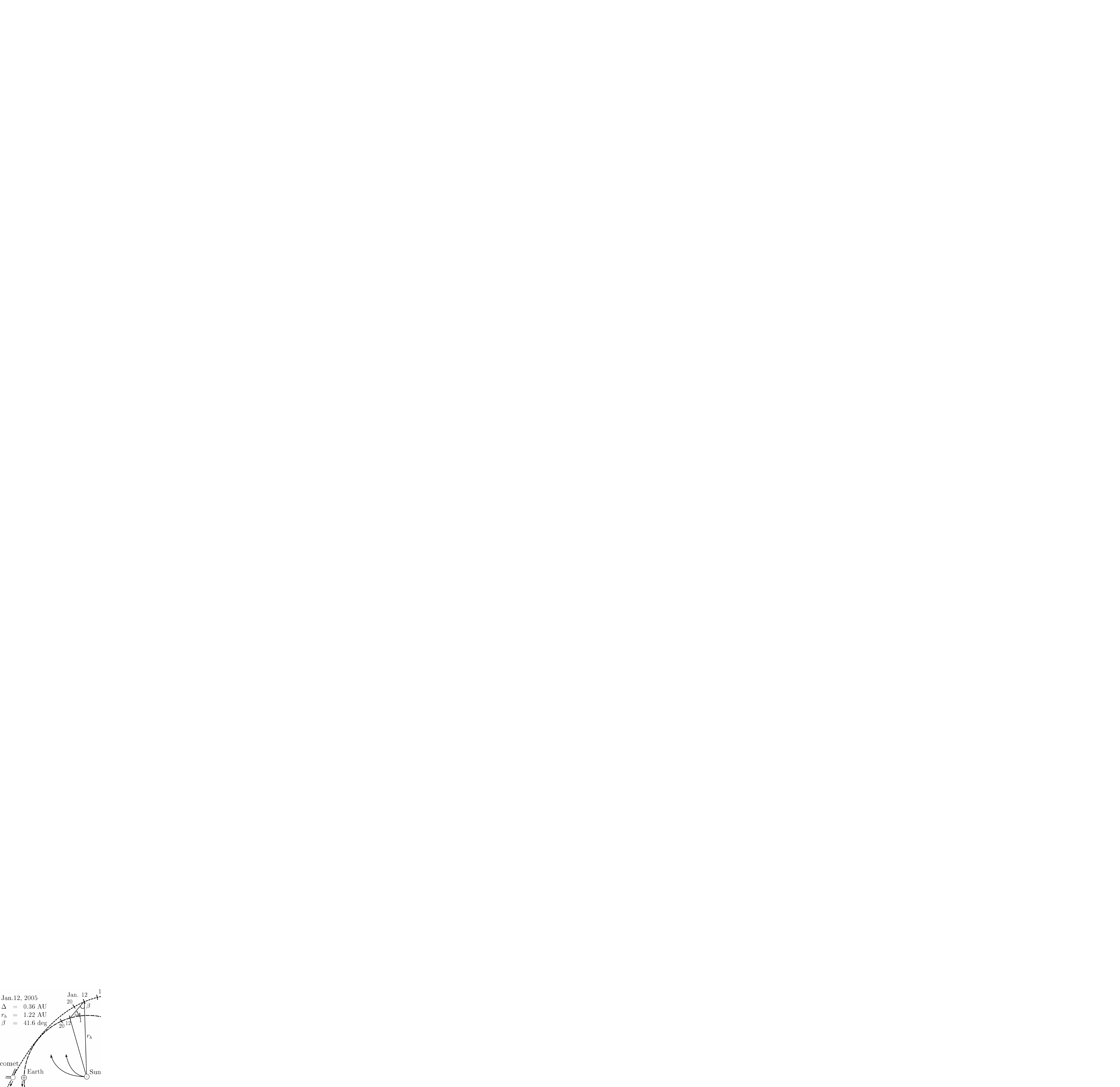}
\caption{Configuration of C/2004 Q2 (Machholz), the Earth, and Sun on 2005 Jan. 12, when the solar wind regime changes suddenly from slow to fast, and ion folding rays are visible. All orbits are seen from above and projected onto the ecliptic. Note that Machholz' orbit has an inclination of $\sim$39\,deg with respect to the ecliptic (data from {\small\ttfamily http://ssd.jpl.nasa.gov/horizons.cgi}).}\label{fig:pdegroote:configuration}
\end{figure}

The velocity of a CIR has both a radial as a corotational component due to the rotation of the Sun. Thus, if we want to extrapolate the SOHO/CELIAS measurements to the position of comet C/2004 Q2 (Machholz) during the observation run, we have to make certain that the comet is not situated too far away from L1, that it is not too high above or below the ecliptic, and that we take both components of the outflow velocity into consideration. The first two conditions are immediately set (Fig. \ref{fig:pdegroote:configuration}). Next, the actual extrapolation is done by using the formula \citep{neugebauer00}
\begin{equation}\Delta t=\Delta t_{rad}+\Delta t_{rot}\ .\label{eq:pdegroote:corotmap}\end{equation}
The first term on the right hand side of equation (\ref{eq:pdegroote:corotmap}) equals $\Delta r/v_\odot$ with $v_\odot$ the velocity of the solar wind and $\Delta r$ the difference between the heliocentric distances of L1 and comet C/2004 Q2 (Machholz). We can set the velocity of the comet $v_k$ to $v_k$\,=\,0 because $v_k$\,$\ll$\,$v_\odot$. The second term on the right hand side can be rewritten as $\Delta t_{rot}$\,=\,$\Delta_{lon}/\Omega_\odot$, with $\Delta_{lon}$ the longitudonal differences between L1 and the comet's position, and $\Omega_\odot$\,=\,2.9\,$\times$\,10$^{-6}$\,rad\,s$^{-1}$ the solar rotation frequency. For example, when the ion rays become visible for the first time, we calculate $\Delta t \sim$6\,h. We emphasise that this method does not provide us with exact values of solar wind parameters at the comet's location, as it does not account for any magnetohydrodynamical phenomena, such as shocks and CMEs. That is why the extrapolation of the SOHO data does not result in a smooth function, but sometimes gives two predictions of the solar wind velocity at the same moment: wind particles travelling faster but originating later than slower particles may, according to our approximations, arrive at the comet's position at exactly the same time, rather than result in a shock front.

Taking these limitations into consideration, we can conclude that the change from a slow to fast solar wind regime, in the form of a corotating interaction region detected on 12 January in L1, is expected to have arrived at C/2004 Q2 (Machholz) on the same date the ion rays are visible in our images.

\section{Discussion}
Overall, there are three possible explanations for the observed features. First, they could be caused by eruptions from the nucleus. This is highly unlikely, because the ion rays are too numerous, too symmetric, always located in the tailward hemisphere, and they change their position too rapidly (e.g. the images taken on 22:50 UT and 23:06 UT, Fig. \ref{fig:pdegroote:foldings_comp}). Second, the tail rays could look like elongations of the shock front. This is not very likely either: considering a gas production rate $Q$ of $2.6\times 10^{29}\mbox{s}^{-1}$ \citep{bonev06}, the shock front is expected to be situated at a distance $R_S>90\times 10^3$ km from the nucleus \citep{wegmann04}, which is just outside the field of view, but the features  are likely to be within the bow shock. A second counterargument is that a shock front is only visible in X-rays \citep{wegmann98} or by in-situ measurements \citep{fuselier91}. A third explanation invokes discontinuities in the solar wind \citep{ip04}. This is the most likely explanation since such a discontinuity was detected by SOHO around the time of the observation. Apparently, a discontinuity in the solar wind, caused by a CIR, is a good candidate for triggering the appearance of cometary folding rays.

\section{Conclusion}
Comet C/2004 Q2 (Machholz) reached a visual brightness of $\sim$3.5\,mag. at its closest approach to Earth on 2005 Jan. 5, making it one of the most spectacular astronomical events of that year. 
The comet was visible during a significant part of the night, making it ideally suited to performing a dedicated ground-based monitoring. The first objective of our intensive monitoring 
was to derive a rotation period \citep{reyniers08}. In addition to this, coma structures were also studied in three different bands. During one night, 12 January, clear ion rays were detected in the U band images after enhancement. Due to the comet's position relatively close to the Lagrangian point L1 and close to the ecliptic, we were able to extrapolate the solar wind velocity data from SOHO to the comet's position. It turned out that the development of the ion rays coincide perfectly  with a sudden change in the local solar wind from the ``slow'' ($\sim$\,400\,km\,s$^{-1}$) to the ``fast'' ($\sim$\,700\,km\,s$^{-1}$) regime. To our knowledge, this is the first time that the relatively rare event of folding ion rays in the coma of a comet could be correlated with a change of the in-situ solar wind state.

\begin{acknowledgements}
The authors want to thank Christoffel Waelkens for the splendid 
observations, and the Mercator staff at La Palma for their indispensable 
support. This research has made use of SOHO/CELIAS and ACE/SWICS data, and 
of the HORIZONS ephemeris service operated at JPL. MR acknowledges
financial support from the Fund for Scientific Research - Flanders 
(Belgium).
\end{acknowledgements}

\bibliographystyle{aa}
\bibliography{komeetlttr3}

\end{document}